\documentclass[a4paper]{jpconf}
\usepackage{graphicx}
\usepackage[latin1]{inputenc}

\begin{document}

\title{Field Theory Renormalization Group: The Tomonaga-Luttinger Model
Revisited}
\author{A.Ferraz}
\address{International Centre for Condensed Matter Physics,
Universidade de Brasilia, 70904-910 Brasilia-DF, Brazil}

\ead{aferraz@unb.br}

\begin{abstract}
We apply field theoretical renormalization group (RG) methods to
describe the Tomonaga-Luttinger model as an important test ground to
deal with spin-charge separation effects in higher spatial
dimensions. We calculate the RG equations for the renormalized
forward couplings $g_{2R}$ and $g_{4R}$ up to two-loop order and
demonstrate that they don't flow in the vicinities of the Fermi
points (FPs). We solve the Callan-Symanzik equation for
$G_{\left(a\right)R}$ in the vicinities of the FPs. We calculate the
related spectral function and the momentum distribution function at
$p=k_{F},p_{0}=\omega$. We compute the renormalized one-particle
irreducible function
$\Gamma_{\left(+\right)R}^{\left(2\right)}\left(p,p_{0}=0;\Lambda\right)$
and show it carries important spin-charge separation effects in
agreement with well known results. Finally we discuss the
implementation of the RG scheme taking into account the important
simplifications produced by the Ward identities.
\end{abstract}

\section{Introduction}

The discovery of high-temperature superconductors and other related
oxide based materials have attracted a lot of attention regarding
the unique properties of low dimensional solids. The simplest and
more notorious representation of a one-dimensional (1d) conductor is
the the so-called Tomonaga-Luttinger (TL) model\cite{Tomonaga}. That
model essentially describes 1d electrons in the vicinity of their
right and left Fermi points interacting with each other by means of
forward scattering processes. Due to the special equivalence
relations between bosons and fermions and left and right charge
conservation laws the TL model can be solved exactly by either
bosonization\cite{Mattis} or, making full use of the Ward
identities, by quantum field theory methods\cite{Larkin}.

The one-particle Green's function for the TL model can also be
calculated exactly by those methods. However such a direct solution
is provided only in coordinate space. To obtain the associated
spectral function which determines the photoemission and the inverse
photoemission spectra one needs to evaluate an intricate double
Fourier transform and perform numerical
approximations\cite{Schonhammer}.

Bosonization methods are hard to implement in higher
dimensions\cite{Haldane}. In addition, those Ward identities
simplifications are inherent to 1d and to the exact number
conservation of right (+) and left (-) particles. As a result to
deal with strongly correlated fermions in two or more spatial
dimensions one generally resorts to other schemes. One of those
alternative approaches is the renormalization group (RG) method. RG
methods were used with success to predict the various instabilities
of the 1d electron gas as a function of the existing coupling
parameters\cite{Solyom}. The RG flows of the renormalized couplings
are also in agreement with the bosonization solutions in several
related problems. Despite that the RG method is not frequently used
to perform a direct calculation of the single-particle Green's
function and its associated spectral function\cite{Ferraz,Busche}.
Since in the TL model there are only low-energy bosonic spin and
charge collective modes it is not clear that the right physics will
emerge out of approximate fermionic RG schemes. Besides, it is also
fair to say that RG methods have not yet been proved capable of
dealing, in full force, with spin-charge separation effects. Within
the g-ology framework the role of the forward coupling $g_{4}$ has
not been fully explored in the RG context so far. One should
therefore try to establish how to deal with spin-charge effects in
well known grounds, such as the TL model, before implementing RG
methods in  more difficult problems such as spin-charge separated
states in higher spatial dimensions.

In this work we apply the field theoretical RG method\cite{Ferraz2}
to describe the TL model in detail. This work is not intended to
review exhaustively all the different RG applications for the TLM.
We refer the reader to several authoritative review articles for
that purpose\cite{Metzner}. Rather we concentrate here uniquely on
the field theoretical RG description of the TLM. Using this scheme
we derive the flow equations for the forward couplings $g_{2R}$ and
$g_{4R}$ up to two-loop order. In calculating the respective
renormalized one-particle irreducible functions $\Gamma_{2R}^{(4)}$
and $\Gamma_{4R}^{(4)}$ we show that non-parquet vertex
contributions are the only source of logarithmic divergences in our
perturbative calculations. Taking into account the self-energy
corrections calculated earlier we demonstrate that, in two-loop
order, the non-flow condition for $g_{2R}$ and $g_{4R}$ is assured
by the exact cancelation produced by the contributions originated by
the anomalous dimension and by the counterterms added to our
renormalized TL Lagrangian. Based on our perturbative RG results we
construct and solve the Callan-Symanzik equation for the
renormalized one-particle Green's function $G_{R}$ at the Fermi
points. We show that it correctly possesses a branch cut structure
due to its non trivial anomalous dimension. Using $G_{R}$ we derive
the spectral function and the momentum distribution function at the
Fermi points. Our results are in agreement with earlier bosonization
work\cite{Schonhammer} as well as more recent RG
estimates\cite{Ferraz2}. Using a momentum RG scale parameter we use
perturbation theory to derive $\Gamma_{R}^{(2)}=G_{R}^{-1}$ at the
Fermi energy $p_{0}=0$ and in the vicinity of $p=\pm k_{F}$. One
important feature emerges naturally from our result. There are now
two emergent characteristic velocities which we relate immediately
to the spin-charge separation effects. Moreover, the general form of
the resulting $\Gamma_{R}^{(2)}$ is in qualitative agreement with
the Dzyaloshinskii and Larkin's Ward identity
solution(DL)\cite{Larkin}. Our estimates are entirely perturbative
since at that stage we only consider contributions up to two-loop
order. To go beyond perturbation theory we relate the effective
two-particle propagators $D_{+-}$ and $D_{++}$, introduced earlier
by DL, to $\Gamma_{2R}^{(4)}$ and $\Gamma_{4R}^{(4)}$ and show how
to implement the regularization, up to infinite order, of all the
divergences which are originated in the perturbation series for
these two one-particle irreducible functions.

Recently new versions of the functional RG containing both fermionic
and auxiliary bosonic fields have been developed with
success\cite{Kopietz,Wetterich}. In one of those\cite{Kopietz}, the
Ward identities are also used to truncate the vertex functions
hierarchical equations and as a result they are able to reproduce
the correct TL spectral function at the Fermi points for the case in
which $g{}_{4}=0$.

\section{The TL Lagrangian Model}

Following the so-called g-ology convention the renormalized TL
fermionic lagrangian is given by

\begin{eqnarray}
&&L_{TL}=\sum_{p,a=\pm,\sigma}(1+\Delta
Z)\psi_{R(a)\sigma}^{\dagger}(p,t)(i\partial_{t}-v_{F}(\left|p\right|-k_{F}))\psi_{R(a)\sigma}(p,t)\nonumber
\\&&-\sum_{p_{1}p_{2}p_{3} \atop \sigma_{1},\sigma_{2}}(g_{2R}+\Delta
g_{2R})\psi_{R(+)\sigma_{1}}^{\dagger}(p_{3},t)\psi_{R(-)\sigma_{2}}^{\dagger}(p_{1}+p_{2}-p_{3},t)\psi_{R(-)\sigma_{2}}(p_{2},t)\psi_{R(+)\sigma_{1}}(p_{1},t)\nonumber
\\&&-\frac{1}{2}\sum_{p_{1}p_{2}p_{3}\atop
\sigma_{1},\sigma_{2}, a}(g_{4R}+\Delta
g_{4R})\psi_{R(a)\sigma_{1}}^{\dagger}(p_{3},t)\psi_{R(a)\sigma_{2}}^{\dagger}(p_{1}+p_{2}-p_{3},t)\psi_{R(a)\sigma_{2}}(p_{2},t)\psi_{R(a)\sigma_{1}}(p_{1},t),\label{lag}
\end{eqnarray}

\begin{figure}[t]
  \centering
  \includegraphics[height=1.0in]{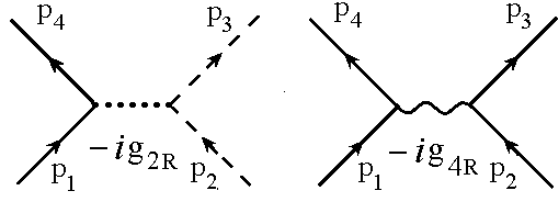}\\
  \caption{The direct renormalized couplings.}\label{inter}
\end{figure}

\noindent where the momentum $p$ is restricted to the interval
$-\lambda\leq\left|p\right|-k_{F}\leq\lambda$ in the vicinities of
the right and left Fermi points and $\Delta Z=Z-1$. In Figure
\ref{inter} we display the direct couplings $g_{2R}$ and $g_{4R}$.
For practical purposes we can also take the limit
$\lambda\rightarrow\infty$. We assume beforehand that neither the
Fermi velocity nor the Fermi momentum are renormalized by
interactions. The quasiparticle weight Z is nullified by
interactions at the Fermi points since there are no stable
quasiparticles in the TL regime but we assume tactically that we
don't know this a priori. In general Z is the multiplicative factor
which relate the bare and the renormalized fermion fields to each
other. The counterterms associated with the contributions from
$\Delta g_{2R}$ and $\Delta g_{4R}$ are included to regularize the
logarithmic divergences which arise in perturbation theory.

The corresponding right and left non-interacting Green's functions
of this model are simply

\begin{equation}
iG_{\left(+\right)}^{\left(0\right)}\left(p,p_{0}\right)=i\left[\frac{\theta\left(p-k_{F}\right)}{p_{0}-v_{F}\left(p-k_{F}\right)+i\delta}+\frac{\theta\left(k_{F}-p\right)}{p_{0}-v_{F}\left(p-k_{F}\right)-i\delta}\right]\label{green1}
\end{equation}

\noindent and

\begin{equation}
iG_{\left(-\right)}^{\left(0\right)}\left(p,p_{0}\right)=i\left[\frac{\theta\left(-\left(p+k_{F}\right)\right)}{p_{0}+v_{F}\left(p+k_{F}\right)+i\delta}+\frac{\theta\left(\left(p+k_{F}\right)\right)}{p_{0}+v_{F}\left(p+k_{F}\right)-i\delta}\right]\label{green2}
\end{equation}

\begin{figure}[h]
  \centering
  \includegraphics[height=0.8in]{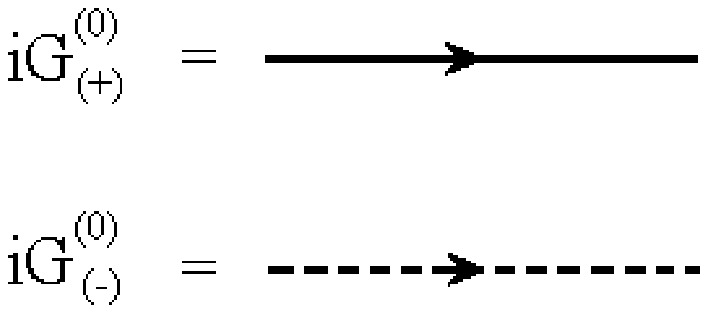}\\
  \caption{The non-interacting Green's functions.}\label{fung}
\end{figure}

Using the appropriate Feynman rules for $L_{TL}$ we can now proceed
with the perturbative calculations of the one-particle irreducible
functions $\Gamma_{2R}^{\left(4\right)}$,
$\Gamma_{4R}^{\left(4\right)}$ and $\Gamma_{2R}^{\left(2\right)}$.
In general, following the renormalization theory, we can relate
those functions at the Fermi points to observable physical
quantities.

Due to the fact that only forward scattering processes are accounted
for in $L_{TL}$ the right $\left(+\right)$ and left $\left(-\right)$
lines never mix in a single loop. As a result in this way no
divergence arises from non-interacting particle-hole loop diagrams.
Taking into account the spin summation factors we find instead

\begin{equation}
\Pi_{\left(+\right)}^{\left(0\right)}\left(q,q_{0}\right)=\frac{iq}{\pi}\left[\frac{\theta\left(q\right)}{q_{0}-v_{F}q+i\delta}+\frac{\theta\left(-q\right)}{q_{0}-v_{F}q-i\delta}\right]\label{pi1}
\end{equation}

\noindent and

\begin{equation}
\Pi_{\left(-\right)}^{\left(0\right)}\left(q,q_{0}\right)=-\frac{iq}{\pi}\left[\frac{\theta\left(-q\right)}{q_{0}+v_{F}q+i\delta}+\frac{\theta\left(q\right)}{q_{0}+v_{F}q-i\delta}\right]\label{pi2}
\end{equation}

These $\Pi^{\left(0\right)}$'s reduce simply to $-i/\pi v_{F}$ at
the Fermi points if we take the limits $q_{0}\rightarrow0$ and
$q\rightarrow0$.

\begin{figure}[h]
  \centering
  \includegraphics[height=1.5in]{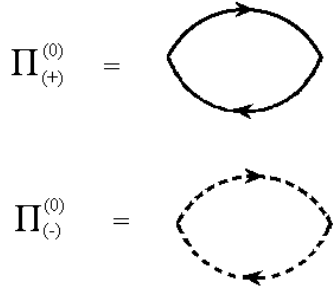}\\
  \caption{The non-interacting particle-hole bubbles.}\label{pi}
\end{figure}

Another important feature due to the neglect of backscattering
processes is that the symmetrized sums of all diagrams containing
closed loops with more than two fermion lines vanish
exactly\cite{Larkin}. Thanks to this important simplification there
are only divergent diagrams in the perturbative series expansions
for both $\Gamma_{2R}^{\left(4\right)}$ and
$\Gamma_{4R}^{\left(4\right)}$ in two-loop order. Those divergences,
which are precisely the object of attention of renormalization
theory, are cancelled out exactly by the counterterms added to the
Lagrangian model. However at this order of perturbation theory we
are also obliged to take into account self-energy corrections.
Therefore before discussing the RG flow of the renormalized
couplings we apply the RG method for the calculation of the
self-energy $\Sigma_{R\left(+\right)}$.

\section{One-Particle Irreducible Function $\Gamma_{2R}^{\left(2\right)}$
up to Two-Loop Order}

Computing diagrammatically the self-energy
$\Sigma_{R\left(+\right)}\left(p,p_{0}\right)$ for the TL model up
to two-loop order, we find:

\begin{figure}[h]
  \centering
  \includegraphics[height=1.3in]{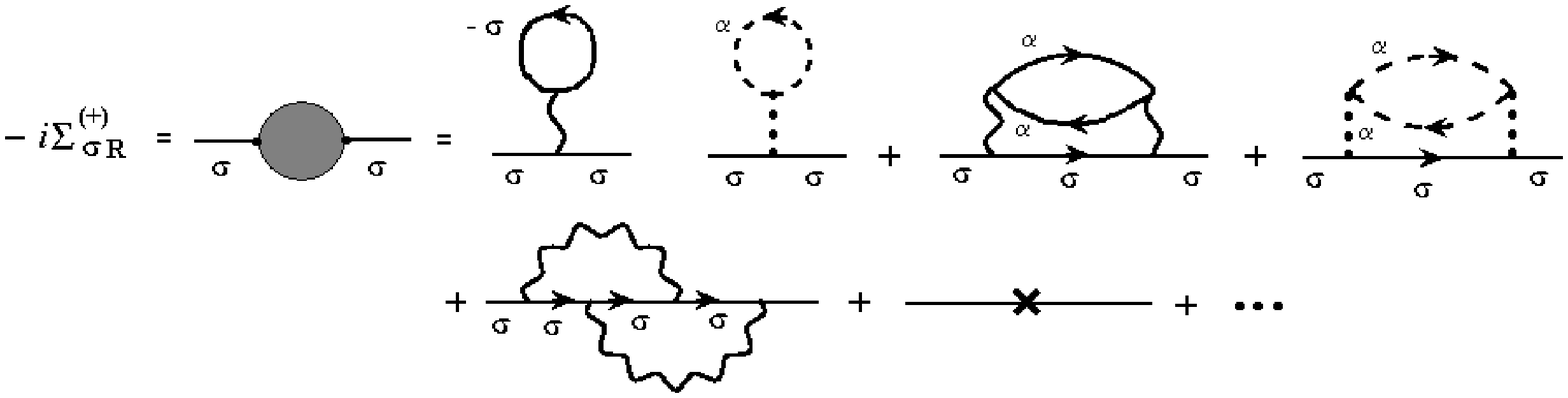}\\
  \caption{The self-energy $\Sigma_{R(+)\sigma}$.}\label{selfener}
\end{figure}

Evaluating these diagrams we obtain

\begin{equation}
\Sigma_{R\left(+\right)}\left(p,p_{0}\right)=\Sigma_{R\left(+\right)}^{\left(a\right)}\left(p,p_{0}\right)+\Sigma_{R\left(+\right)}^{\left(b\right)}\left(p,p_{0}\right)+\Sigma_{R\left(+\right)}^{\left(c\right)}\left(p,p_{0}\right),\label{d1}
\end{equation}

\noindent where

\begin{equation}
\Sigma_{R\left(+\right)}^{\left(a\right)}\left(p,p_{0}\right)=\frac{\overline{g}_{2R}\Omega}{2\pi}+\frac{\overline{g}_{4R}\Omega}{4\pi}+\left(\frac{\overline{g}_{4R}}{2\pi\sqrt{2}}\right)^{2}\left(v_{F}\left(p-k_{F}\right)\right)^{2}G_{\left(+\right)}^{\left(0\right)}\left(p,p_{0}\right),\label{d2}
\end{equation}

\begin{eqnarray}
\Sigma_{R\left(+\right)}^{\left(b\right)}\left(p,p_{0}\right) & = &
-\Delta Z\left(p_{0}-v_{F}\left(p-k_{F}\right)\right)\label{d3}
\end{eqnarray}

\noindent and

\begin{eqnarray}
\Sigma_{R\left(+\right)}^{\left(c\right)}\left(p,p_{0}\right)=-\left(\frac{\overline{g}_{2R}}{2\pi\sqrt{2}}\right)^{2}\left(p_{0}-v_{F}\left(p-k_{F}\right)\right)\left\{ \ln\left(\frac{\Omega-v_{F}\left|\Delta p\right|+p_{0}-i\delta}{v_{F}\left|\Delta p\right|+p_{0}-i\delta}\right)\right.\nonumber \\
\left.+\ln\left(\frac{\Omega-v_{F}\left|\Delta
p\right|+p_{0}-i\delta}{v_{F}\left|\Delta
p\right|+p_{0}-i\delta}\right)\right\} \label{d4}
\end{eqnarray}

\noindent where $\Delta p=p-k_{F}$,
$\overline{g}_{iR}=\overline{g}_{iR}/v_{F}$ and
$\Omega=2v_{F}\lambda$.

Having established what $\Sigma_{R\left(+\right)}$ is we move on to
calculate the related one-particle Green's function
$G_{\left(+\right)R}=\left(\Gamma_{R\left(+\right)}^{\left(2\right)}\right)^{-1}$
where

\begin{equation}
\Gamma_{R\left(+\right)}^{\left(2\right)}\left(p,p_{0}\right)=p_{0}-v_{F}\left(p-k_{F}\right)-\Sigma_{R\left(+\right)}\left(p,p_{0}\right)\label{gamma1}
\end{equation}

To determine $\Delta Z\left(\omega\right)$ we define
$\Gamma_{R\left(+\right)}^{\left(2\right)}$ such that, at the RG
scale $p_{0}=\omega$, and at the Fermi point $\Delta p=0$,
$\Gamma_{R\left(+\right)}^{\left(2\right)}\left(p=k_{F},p_{0}=\omega;\omega\right)=\omega$.
Using our two-loop results we get

\begin{equation}
Z\left(\omega\right)=1-\left(\frac{\overline{g}_{2R}}{2\pi}\right)^{2}\ln\left(\frac{\Omega}{\omega}\right)\label{z1}
\end{equation}

Notice that the quasiparticle weight $Z\left(\omega\right)$
satisfies the RG equation

\begin{equation}
\omega\frac{d\ln Z\left(\omega\right)}{d\omega}=\gamma,\label{zr}
\end{equation}

\noindent where $\gamma$ is the anomalous dimension given by

\begin{equation}
\gamma=\left(\frac{\overline{g}_{2R}}{2\pi}\right)^{2}\label{gammar}
\end{equation}

In view of this $Z\left(\omega\right)$ scales simply with the RG
parameter as

\begin{equation}
Z\left(\omega\right)=\left(\frac{\omega}{\Omega}\right)^{\gamma},\label{zr1}
\end{equation}

\noindent and since $\gamma\geq0$ for non-zero forward coupling
$Z\left(\omega\right)$ is nullified at the FP when
$\omega\rightarrow0$ in agreement with the fact that there are no
well defined quasiparticles in the TL state. This result was
obtained earlier in Refs. \cite{Metzner,Kimura}.

\section{Flow Equations for the Renormalized Couplings}

Following the renormalization theory we can relate the physical
couplings $g_{2R}$ and $g_{4R}$ to the-particle irreducible
functions $\Gamma_{2R}^{\left(4\right)}$ and
$\Gamma_{4R}^{\left(4\right)}$ at the Fermi points. In this way for
the external momenta $p_{1}=p_{3}=k_{F}$, $p_{2}=-k_{F}$ and
$p_{10}=p_{20}=p_{30}=\omega/2\approx0$ we define

\begin{equation}
\Gamma_{2R}^{\left(4\right)}\left(p_{1}=k_{F},p_{2}=-k_{F};p_{1}+p_{2}-p_{3}=-k_{F},p_{3}=k_{F};p_{10}=p_{20}=p_{30}=\omega/2;\omega\right)=-ig_{2R}\left(\omega\right)\label{gamma2}
\end{equation}

\noindent and similarly, for $p_{1}=p_{3}=p_{2}=k_{F}$ and
$p_{10}=p_{20}=p_{30}=\omega/2$,

\begin{equation}
\Gamma_{4R}^{\left(4\right)}\left(p_{1}=k_{F},p_{2}=k_{F};p_{1}+p_{2}-p_{3}=k_{F},p_{3}=k_{F};p_{10}=p_{20}=p_{30}=\omega/2;\omega\right)=-ig_{4R}\left(\omega\right)\label{gamma3}
\end{equation}

\begin{figure}[t]
  \centering
  \includegraphics[height=2.3in]{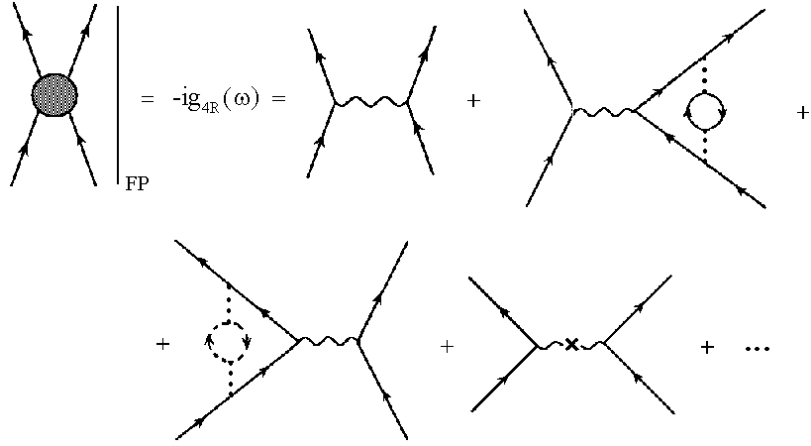}\\
  \caption{The $g_{4R}$ channel.}\label{g4r}
\end{figure}

Using standard diagrammatic analysis as shown in the Figure
\ref{g4r} we have at the vicinity of the Fermi point

\begin{equation}
\Gamma_{4R}^{\left(4\right)}=-ig_{4R}\left(\omega\right)-i\frac{g_{4R}\left(\omega\right)\overline{g}_{2R}^{2}\left(\omega\right)}{2\pi^{2}}\ln\left(\frac{\Omega}{\omega}\right)-i\Delta
g_{4R}\left(\omega\right)+...\label{gamma4}
\end{equation}

To be consistent with the definition given above in eqn.\ref{gamma3}
we then set

\begin{equation}
\Delta
g_{4R}\left(\omega\right)=-\frac{g_{4R}\left(\omega\right)\overline{g}_{2R}^{2}\left(\omega\right)}{2\pi^{2}}\ln\left(\frac{\Omega}{\omega}\right)\label{counter1}
\end{equation}

\begin{figure}[b]
  \centering
  \includegraphics[height=2.3in]{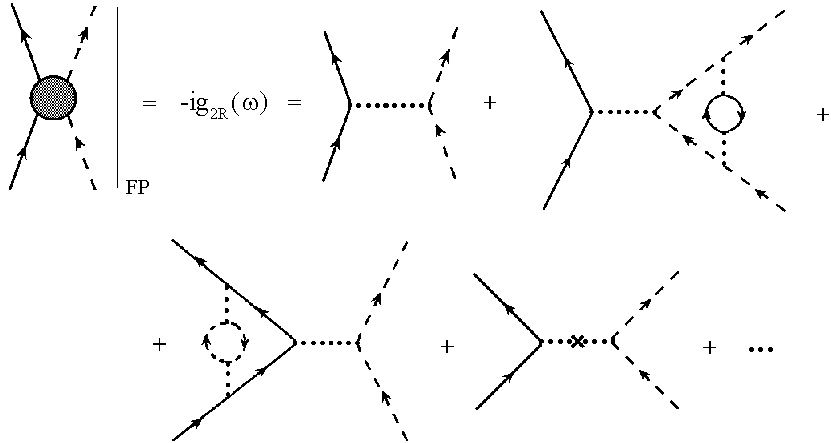}\\
  \caption{The $g_{2R}$ channel.}\label{g2rr}
\end{figure}

\begin{figure}[t]
  \centering
  \includegraphics[height=1.1in]{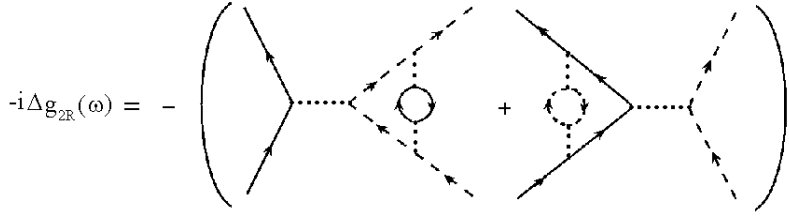}\\
  \caption{The counterterm $-i\Delta g_{2R}(\omega)$.}\label{cg2r}
\end{figure}

We display the diagrams for $\Gamma_{2R}^{(4)}|_{FP}$ in Figure
\ref{g2rr}. Following the same strategy as before we indicate in
Figure \ref{cg2r} how to proceed with the calculation of the
counterterm $-i\Delta g_{2R}(\omega)$.

From RG theory we know that the bare and renormalized couplings are
related to each other by

\begin{equation}
g_{iB}=Z^{-2}\left(\omega\right)\left(g_{iR}+\Delta
g_{iR}\right)\label{gir}
\end{equation}

Since the bare couplings are independent of the RG scale $\omega$
they satisfy automatically the RG conditions

\begin{equation}
\omega\frac{dg_{2B}}{d\omega}=\omega\frac{dg_{2B}}{d\omega}=0\label{g2r}
\end{equation}

The flow equations for the corresponding renormalized couplings
follow from these conditions. We find immediately that

\begin{equation}
\beta_{i}(\omega)=\omega\frac{dg_{iR}}{d\omega}=2\gamma
g_{iR}-\omega\frac{d\Delta g_{iR}}{d\omega}\label{gir2}
\end{equation}

Using our perturbative results for $\gamma$, $\Delta g_{2R}$ and
$\Delta g_{4R}$ we then get that these flow equations are exactly
nullified for the TL liquid:

\begin{equation}
\beta_{2}(\omega)=\omega\frac{dg_{2R}}{d\omega}=\beta_{4}(\omega)=\omega\frac{dg_{4R}}{d\omega}=0\label{g2r2}
\end{equation}

A rigorous mathematical proof of the vanishing of the beta functions
in all orders of perturbation theory in the TLM is given in Ref.
\cite{Benfatto}.

\section{RG Calculation of the One-Particle Green's Function at the Fermi
Points}

As we saw in the previous section the TL liquid is characterized by
the non-flow of the renormalized forward couplings. This simplifies
the RG approach to this problem. In this scheme we relate, quite
generically, the renormalized Green's function $G_{R}$ to its bare
analogue $G_{B}$ by

\begin{equation}
G_{\left(a\right)R}\left(p,p_{0};g_{2R};g_{4R};\omega\right)=Z^{-1}\left(\omega\right)G_{\left(a\right)B}\left(p,p_{0};g_{2B};g_{4B}\right)\label{greena}
\end{equation}

Since $G_{B}$ is independent of the RG scale $\omega$ we derive the
Callan-Symanzik equation (CZE)\cite{Perskin} differentiating
$G_{\left(a\right)B}$ with respect to $\omega$. Taking into account
that the $g_{iR}$'s don't flow , in the vicinity of the Fermi
points, for $p=\pm k_{F}$, the CZE reduces to

\begin{equation}
\left(\omega\frac{\partial}{\partial\omega}+\gamma\right)G_{\left(a\right)R}\left(p=\pm
k_{F},p_{0};\omega\right)=0\label{greenb}
\end{equation}

Since $G_{\left(a\right)R}$ has ordinary dimension of
$\left(energy\right)^{-1}$, on dimensional grounds we have that

\begin{equation}
\left(\omega\frac{\partial}{\partial\omega}+p_{0}\frac{\partial}{\partial
p_{0}}\right)G_{\left(a\right)R}\left(p=\pm
k_{F},p_{0};\omega\right)=-G_{\left(a\right)R}\left(p=\pm
k_{F},p_{0};\omega\right)\label{greenc}
\end{equation}

We can use this to rewrite the CZE in the form

\begin{equation}
\left(p_{0}\frac{\partial}{\partial
p_{0}}+1-\gamma\right)G_{\left(a\right)R}\left(p=\pm
k_{F},p_{0};\omega\right)=0\label{greend}
\end{equation}

Considering that the anomalous dimension $\gamma$ is constant, this
equation can be easily integrated to give

\begin{equation}
G_{\left(a\right)R}\left(p=\pm
k_{F},p_{0};\omega\right)=\frac{1}{\omega}\left(\frac{\omega}{p_{0}+i\delta}\right)^{1-\gamma}\label{greenf}
\end{equation}

\noindent which has indeed a branch cut structure as opposed to the
simple pole nature, typical of the Fermi liquid one-particle Green's
function. Consequently, for $p_{0}\leq0$, at the Fermi points,

\begin{equation}
G_{\left(a\right)R}\left(p=\pm
k_{F},p_{0};\omega\right)=-\frac{1}{\omega}\left(\frac{\omega}{\left|p_{0}\right|}\right)^{1-\gamma}\left(\cos\pi\gamma+i\sin\pi\gamma\right)\label{greeng}
\end{equation}

It follows from this that the spectral function
$A_{\left(a\right)R}\left(p=\pm k_{F},p_{0};\omega\right)$ reduces
to

\begin{equation}
A_{\left(a\right)R}\left(p=\pm
k_{F},p_{0};\omega\right)=\theta\left(-p_{0}\right)\left(\frac{\left|p_{0}\right|}{\omega}\right)^{\gamma}\frac{\sin\pi\gamma}{\left|p_{0}\right|}\label{greenh}
\end{equation}

This result is in agreement with earlier RG estimates of Kopietz and
coworkers\cite{Busche}. To calculate next the momentum distribution
function $n\left(p=\pm k_{F}\right)$ at the FPs, it suffices to
integrate $A_{\left(a\right)R}$ over an energy interval of width
$2\omega$ around $p_{0}=0$. We find

\begin{equation}
n\left(p=\pm
k_{F}\right)=\frac{\sin\pi\gamma}{2\pi\gamma}\label{greeni}
\end{equation}

\noindent which reduces to $n\left(\pm k_{F}\right)=1/2$ in the
limit $\gamma\rightarrow0$.

Let us now consider the alternative limit $p_{0}=0$ and $\Delta
p=\Lambda\approx0$ where the parameter $\Lambda$ is now our RG
momentum scale which will be used to take the physical system
towards the FPs. If we repeat our RG analysis for
$\Gamma_{iR}^{\left(4\right)}$ making use of such RG scale we find
at the FPs

\begin{equation}
\Delta
g_{iR}\left(\omega\right)=-\frac{g_{iR}\left(\omega\right)\overline{g}_{2R}^{2}\left(\omega\right)}{2\pi^{2}}\ln\left(\frac{2\lambda}{\Lambda}\right)\label{greenj}
\end{equation}

Since the forward couplings don't flow in the TL we can determine
the corresponding $Z\left(\Lambda\right)$ taking into account that
we must have as before

\begin{equation}
2\gamma_{\Lambda}g_{iR}=\Lambda\frac{d\Delta
g_{iR}}{d\Lambda}=\frac{g_{iR}\left(\omega\right)\overline{g}_{2R}^{2}\left(\omega\right)}{2\pi^{2}}\label{greenk}
\end{equation}

\noindent Consequently,

\begin{equation}
\Lambda\frac{d\ln
Z\left(\Lambda\right)}{d\Lambda}=\gamma_{\Lambda}=\gamma\label{greenl}
\end{equation}

\noindent and again we have that

\begin{equation}
Z\left(\Lambda\right)=Z\left(2\lambda\right)\left(\frac{\Lambda}{2\lambda}\right)^{\gamma}\rightarrow0\label{greenm}
\end{equation}

\noindent as $\Lambda\rightarrow0$ for $\gamma>0$, which is
consistent with the fact that there are no quasiparticles at the
FPs.

Invoking once again the RG condition between the bare and
renormalized $\Gamma_{\left(+\right)}^{\left(2\right)}$ 's we write

\begin{equation}
\Gamma_{\left(+\right)R}^{\left(2\right)}\left(p,p_{0}=0;\Lambda\right)=Z\left(\Lambda\right)\Gamma_{\left(+\right)B}^{\left(2\right)}\left(p,p_{0}=0\right)\label{greenn}
\end{equation}

In the weak coupling limit in a regime consistent with our
perturbative two-loop results $Z\left(\Lambda\right)$ reduces to

\begin{equation}
Z\left(\Lambda\right)=1+\gamma\ln\left(\frac{\Lambda}{2\lambda}\right)+...\label{greeno}
\end{equation}

Combining this with our earlier results produces

\begin{equation}
\Gamma_{\left(+\right)B}^{\left(2\right)}\left(p,p_{0}=0\right)=-v_{F}\Delta
p\left[1+\left(\frac{\overline{g}_{2R}}{2\pi}\right)^{2}\ln\left(\frac{2\lambda}{\Delta
p}\right)-\left(\frac{\overline{g}_{4R}}{2\pi\sqrt{2}}\right)^{2}\right]\label{greenp}
\end{equation}

Consequently in the vicinities of the FPs

\begin{equation}
\Gamma_{\left(+\right)R}^{\left(2\right)}\left(p,p_{0}=0;\Lambda\right)=-\frac{u_{c}u_{s}\Delta
p}{v_{F}}\left[1+\left(\frac{g_{2R}}{2\pi\sqrt{u_{c}u_{s}}}\right)^{2}\ln\left(\frac{\Lambda}{\Delta
p}\right)\right],\label{greenq}
\end{equation}

\noindent with

\begin{equation}
u_{c}=v_{F}\left(1+\frac{|\overline{g}_{4R}|}{2\pi\sqrt{2}}\right)\label{greenr}
\end{equation}

\noindent and

\begin{equation}
u_{s}=v_{F}\left(1-\frac{|\overline{g}_{4R}|}{2\pi\sqrt{2}}\right)\label{greens}
\end{equation}

The presence of two different velocities violates the single-pole
character of the one-particle Green's function and it is directly
related to the spin-charge separation which takes place in the
TLM\cite{Voit}. This spin-charge separation is a feature which
emerges naturally from our perturbative RG analysis. Notice that up
to order $O(\overline{g}_{2R}^{2})$, in the weak coupling limit,
$v_{F}\approx\sqrt{u_{c}u_{s}}$ and
$\Gamma_{\left(+\right)R}^{\left(2\right)}$ reduces to

\begin{equation}
\Gamma_{\left(+\right)R}^{\left(2\right)}\left(p,p_{0}=0;\Lambda\right)\approx-\left(u_{c}\Delta
p\right)^{\frac{1}{2}}\left(u_{s}\Delta
p\right)^{\frac{1}{2}}\left(\frac{\Lambda}{\Delta
p}\right)^{\gamma}\label{greent}
\end{equation}

\noindent which is in qualitative agreement with well-known
results\cite{Dzyasloshinskii}. However, our estimates for $u_{c}$
and $u_{s}$ are only valid up to two-loop order. For this reason
they cannot be identical to the velocities obtained from the
solutions of Schwinger-Dyson equations which include contributions
up to infinite order\cite{Larkin}.

\section{Ward Identities}

As demonstrated by Dzyaloshinskii and Larkin\cite{Larkin} the use of
Ward identities greatly simplifies the analysis of the TL model.
Making use of these identities and taking into account the
simplifications produced by the cancelation of the symmetrized sum
of all diagrams containing closed loops with more than two fermion
lines we are naturally led, following DL, to introduce two
interactions propagators $D_{++}$ and $D_{+-}$. These two
propagators can be used to describe the effective two-particle
interactions in the TLM. $D_{++}$ and $D_{+-}$ are finite at the FPs
and do not need to undergo any regularization procedure. In contrast
with that, as we discussed in section 4, the one-particle
irreducible functions $\Gamma_{2R}^{\left(4\right)}$ and
$\Gamma_{4R}^{\left(4\right)}$ produce non-parquet vertex
contributions which need to be regularized order by order in
perturbation series. Therefore to relate the DL interaction
propagators back to the corresponding one-particle irreducible
functions we must include appropriate vertex functions. Considering
these feature we can readily extend our perturbative results writing
the exact one-particle irreducible functions
$\Gamma_{2R}^{\left(4\right)}$ and $\Gamma_{4R}^{\left(4\right)}$ in
the form

\begin{figure}[b]
  \centering
  \includegraphics[width=4.3in]{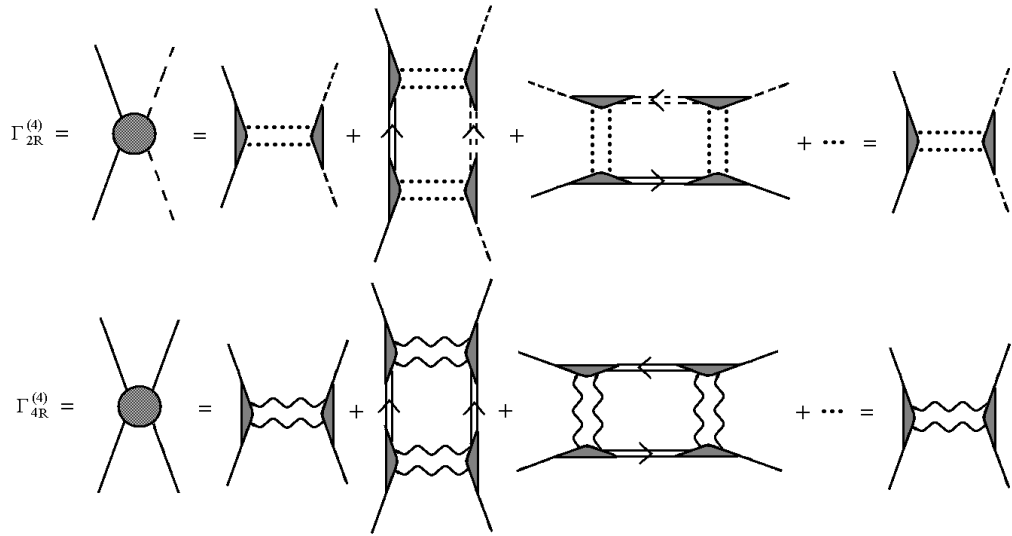}\\
  \caption{The exact one-particle irreducible functions $\Gamma_{2R}^{(4)}$ and $\Gamma_{4R}^{(4)}$.}\label{channels}
\end{figure}

\begin{eqnarray}
\Gamma_{2R}^{(4)}\left(p_{1},p_{2},p_{1}+p_{2}-p_{3},p_{3}\right)=&&\Gamma_{+}^{R}(p_{3},p_{1},p_{3}-p_{1})iD_{+-}\left(p_{1}-p_{3}\right)\nonumber\\
&&.\Gamma_{-}^{R}\left(p_{1}+p_{2}-p_{3},p_{3},p_{1}-p_{3}\right)\label{dyson}
\end{eqnarray}

\noindent and

\begin{eqnarray}
\Gamma_{4R}^{(4)}\left(p_{1},p_{2},p_{1}+p_{2}-p_{3},p_{3}\right)=&&\Gamma_{+}^{R}(p_{1},p_{3},p_{3}-p_{1})iD_{++}\left(p_{1}-p_{3}\right)\nonumber\\
&&.\Gamma_{+}^{R}\left(p_{1}+p_{2}-p_{3},p_{2},p_{1}-p_{3}\right)\label{dyson2}
\end{eqnarray}

\begin{figure}[t]
  \centering
  \includegraphics[height=1.2in]{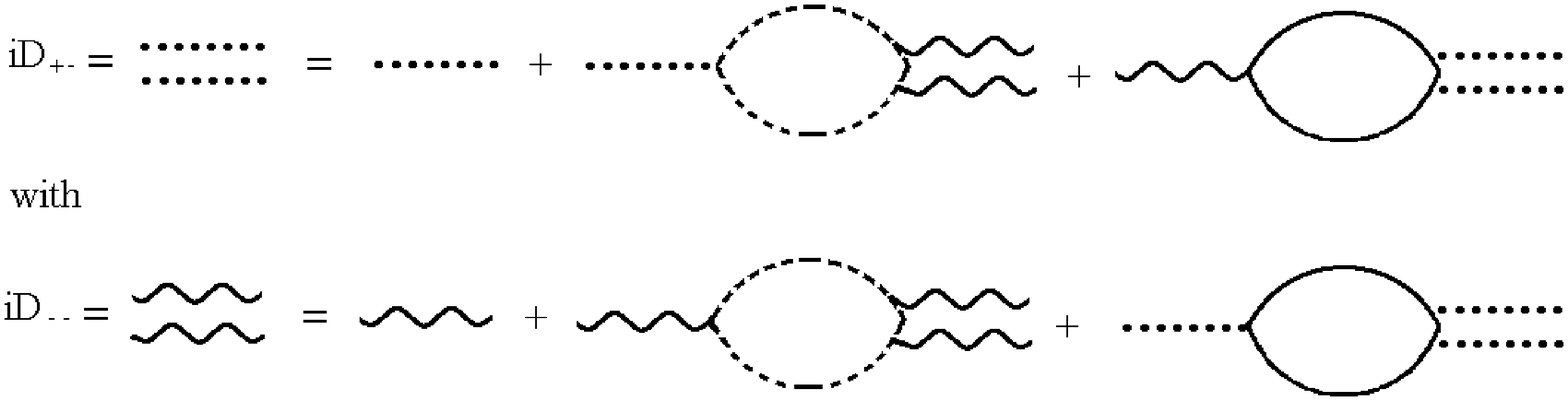}\\
  \caption{The exact interaction propagators $iD_{+-}$ and $iD_{--}$.}\label{dd1}
\end{figure}

\begin{figure}[b]
  \centering
  \includegraphics[height=1.2in]{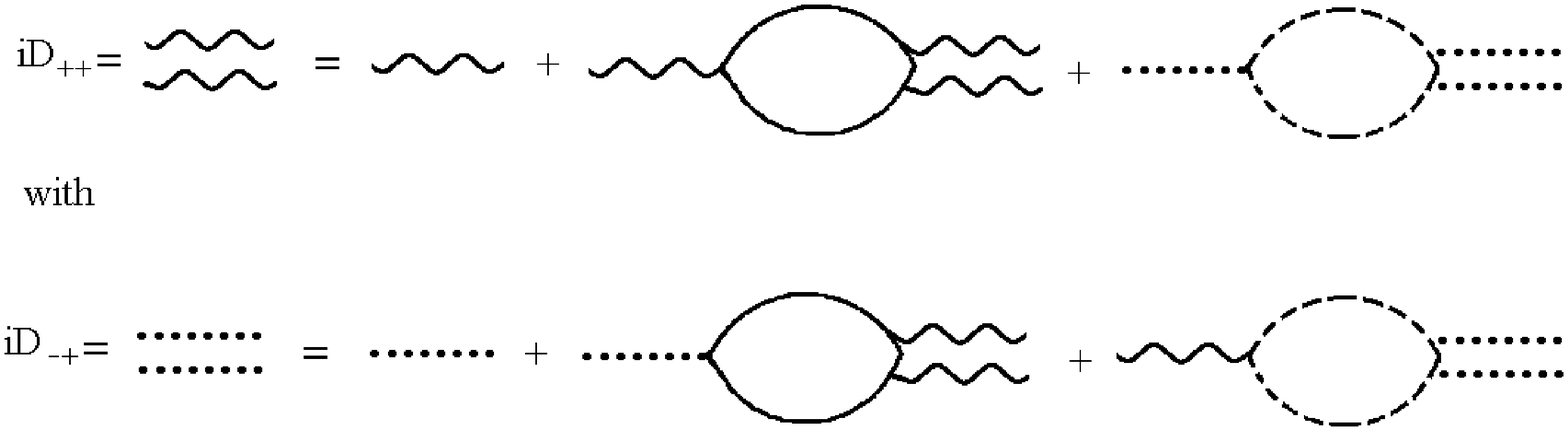}\\
  \caption{The propagators $iD_{++}$ and $iD_{-+}$.}\label{dd2}
\end{figure}

These $\Gamma_{iR}^{(4)}$'s are represented diagrammatically in
Figure \ref{channels}. Thanks to the already mentioned
simplification inherent to the TLM, the propagators $D_{+-}$ and
$D_{++}$ are determined exactly by solving the Schwinger-Dyson
equations indicated in Figures \ref{dd1} and \ref{dd2}. Solving
these SDE's for $D_{++}$ and $D_{+-}$, diagrammatically, we arrive
immediately at the DL results for the charge and spin velocities,
namely,
$u_{c}=v_{F}\sqrt{(1+\overline{g}_{4R}/\pi)^{2}-(\overline{g}_{2R}/\pi)^{2}}$
and $u_{s}=v_{F}$.

\begin{figure}[t]
  \centering
  \includegraphics[height=3.0in]{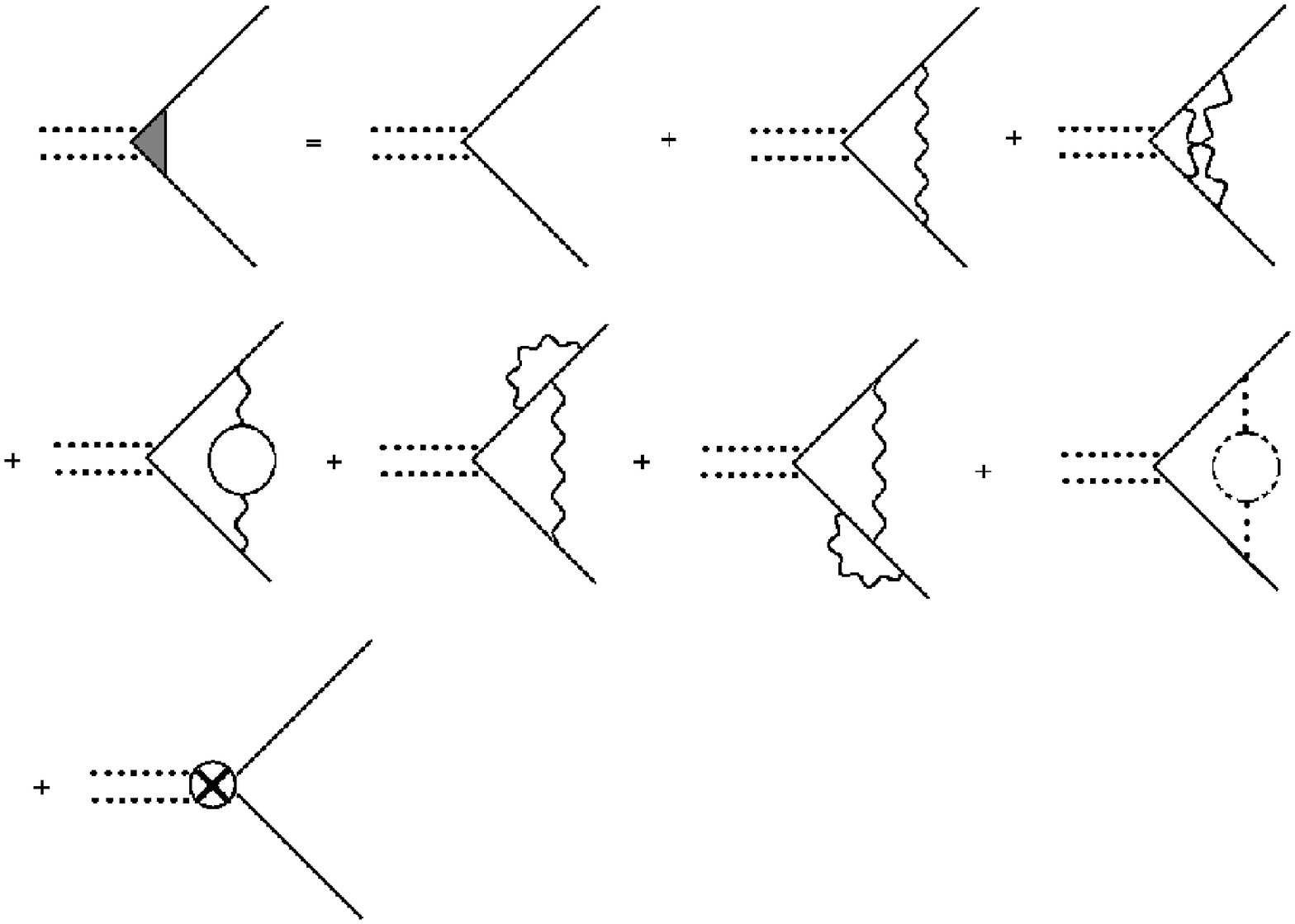}\\
  \caption{The Renormalized vertex function $\Gamma_{+}^{R}$.}\label{suscet}
\end{figure}

Since the interaction propagators are independent of renormalization
parameters they are RG invariants. In contrast the vertex functions
$\Gamma$'s produce logarithmic singularities which need to be
cancelled out by appropriate counterterms. As we showed before, in
the RG framework, we solve this problem either by relating the bare
and renormalized vertices to each other through relations of the
type

\begin{equation}
\Gamma_{a}^{R}=Z_{2}\Gamma_{a}^{B}=\left(1+\Delta
Z_{2}\right)\Gamma_{a}^{B}\label{z2r}
\end{equation}

\noindent with $a=+,-$, or by constructing the necessary counterterm
directly, order by order in perturbation theory. Following this
latter route, up to two-loop order the renormalized vertex function
$\Gamma_{a}^{R}$ is determined by the diagrams shown in Figure
(\ref{suscet}).

\begin{figure}[b]
  \centering
  \includegraphics[height=0.8in]{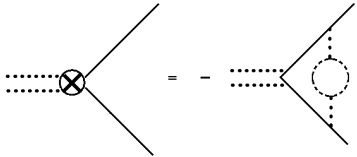}\\
  \caption{The vertex function counterterm $\Delta\Gamma_{+}^{R}$.}\label{countersus}
\end{figure}

If we define $\Gamma_{a}^{R}$ such that
$\left.\Gamma_{a}^{R}\right|_{FP}=1$ it then follows that the vertex
counterterm diagram which appears in Figure (\ref{countersus}) is
determined exactly by this condition.

This gives

\begin{equation}
\left.\Delta\Gamma_{+}^{R}\right|_{FP}=-\left(\frac{\overline{g}_{2R}}{2\pi}\right)^{2}\ln\left(\frac{\Omega}{\omega}\right)=\Delta
Z\label{z2r2}
\end{equation}

It follows from this that the vertex renormalization constant
$Z_{2}$ must be identical to the quasiparticle weight:

\begin{equation}
Z_{2}=Z\label{z2r3}
\end{equation}

Such an equivalence between the vertex renormalization constant
$Z_{2}$ and the quasiparticle amplitude $Z$ was also pointed out in
Ref. \cite{Metzner2}. In fact, since the Ward identity(WI) is
preserved by the renormalization process, this result emerges
naturally from the WI itself. As a consequence of this
simplification, at the FPs, the renormalized one-particle
irreducible functions $\Gamma_{2R}^{\left(4\right)}$ and
$\Gamma_{4R}^{\left(4\right)}$ reduce simply to the interaction
propagators $iD_{+-}$ and $iD_{++}$ respectively, which by being RG
invariants, produce vanishing beta functions to all orders in
perturbation theory in agreement with the rigorous analysis of Ref.
\cite{Benfatto}.

\section{Conclusion}

We implement the field theoretical RG method in the presence of
spin-charge separation effects in one spatial dimension. We take the
Tomonaga-Luttinger model as our reference test ground in dealing
with future higher dimensional problems. As is well known trhe TLM
was solved by bosonization and by quantum field theoretical methods
which take into explicit account the simplifications produced by the
Ward identities. Unfortunately those simplifications are inherent to
one dimension and contrary to the RG scheme both methods are
difficult to implement in more general situations.

Using our RG method we calculate the self-energy up to two-loop
order. From it we find a non-zero anomalous dimension and a
quasiparticle weight which is nullified at the Fermi points. We
compute the RG equations for the renormalized coupling constants
$g{}_{2R}$ and $g{}_{4R}$, up to two-loop order, and demonstrate
that they don't flow in the vicinities of the FPs. We derive the
Callan-Symanzik equation for the one-particle Green's function at
$p=k_{F},p_{0}=\omega$. The CSE is easily integrated producing an
expected branch cut structure in $G_{\left(a\right)R}$. Using this
we derive the spectral function and the momentum distribution
function at $p=k_{F}$. We calculate the one-particle irreducible
function
$\Gamma_{\left(+\right)R}^{\left(2\right)}\left(p,p_{0}=0;\Lambda\right)$
taking into explicit account spin-charge separation effects in the
weak coupling regime. Our results are in qualitative agreement with
the other approaches. Finally we discuss the inclusion of the Ward
identities in the RG scheme. We show that their present forms are
preserved upon the renormalization and thanks to them the
one-particle irreducible functions in the vicinities of the FPs
reduce to interaction propagators which can be solved exactly by
appropriate Schwinger-Dyson equations. To conclude we add that some
of the material discussed in this paper have appeared before in the
literature in one form or another. We tried to relate, whenever
possible, our results to some of those works. By presenting them
here in a field theoretical self-contained form we hope to bring new
insights concerning the RG applications with both spin-charge
separation effects and Ward identities in more general problems.

\pagebreak

\paragraph{Acknowledgements- I wish to acknowledge the discussions and help
I had from Eberth Correa and Hermann Freire in the preparation of
this paper. This work was financially supported by FINEP and the
Min. of Science and Tecnology from Brazil .}

\end{document}